\documentclass{ws-ijmpb}
\usepackage{graphicx}

\begin{document}

\markboth{Julius RANNINGER}
{Cooperative localization-delocalizatino in the high $T_c$ cuprates}

\title{COOPERATIVE LOCALIZATION-DELOCALIZATION IN THE HIGH $T_c$ CUPRATES}

\author{Julius RANNINGER}

\address{Institut N\'eel, CNRS and Universit\'e Joseph Fourier, 25 rue des Martyrs, BP 166\\
38042 Grenoble cedex 9, France\footnote{julius.ranninger@grenoble.cnrs.fr }}

\maketitle

\begin{abstract}
The intrinsic metastable crystal structure of the cuprates  results in local 
dynamical lattice instabilities, strongly coupled to the density fluctuations of the charge carriers. 
They acquire in this way simultaneously both, delocalized and localized features. 
It is responsible for a partial fractioning of the Fermi surface, i.e., the  Fermi surface gets  
hidden in a region around  the anti-nodal points, because of the opening of a pseudogap in the 
normal state, arising from a partial charge  localization. The high energy localized single-particle 
features are a result of a segregation of the homogeneous crystal structure into checker-board 
local nano-size  structures, which breaks the local translational and rotational symmetry. 
The pairing in such a system is dynamical rather than static, whereby charge carriers get 
momentarily trapped into pairs in a deformable dynamically fluctuating ligand environment. 
We conclude  that the intrinsically heterogeneous structure of the cuprates must play an important 
role in this type of superconductivity.
\end{abstract}

\keywords{Localization-delocalization; fermi surface fractioning; local symmetry breaking; crystalline metastability.}

\section{Introduction}

\begin{figure}[h!t]
\begin{minipage}[b]{6cm}
\includegraphics[width=6cm]{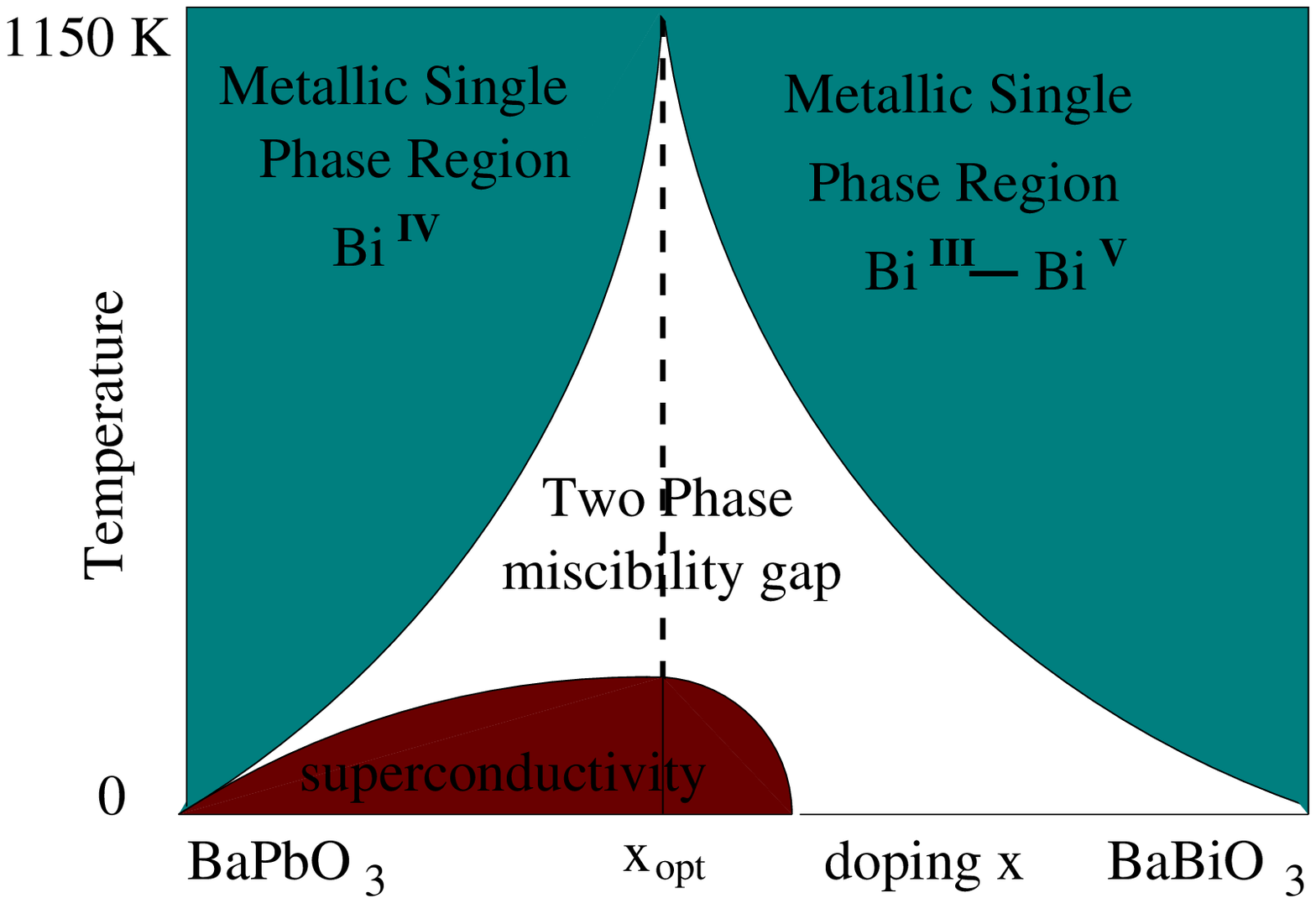}
\caption{Phasediagram for the BaBi$_x$Pb$_{1-x}$O$_3$ synthesization (after Ref.~\protect\refcite{Sleight-1991}).}
\label{Fig1}
\end{minipage}
 \hfill
\begin{minipage}[b]{6cm}
\includegraphics[width=6cm]{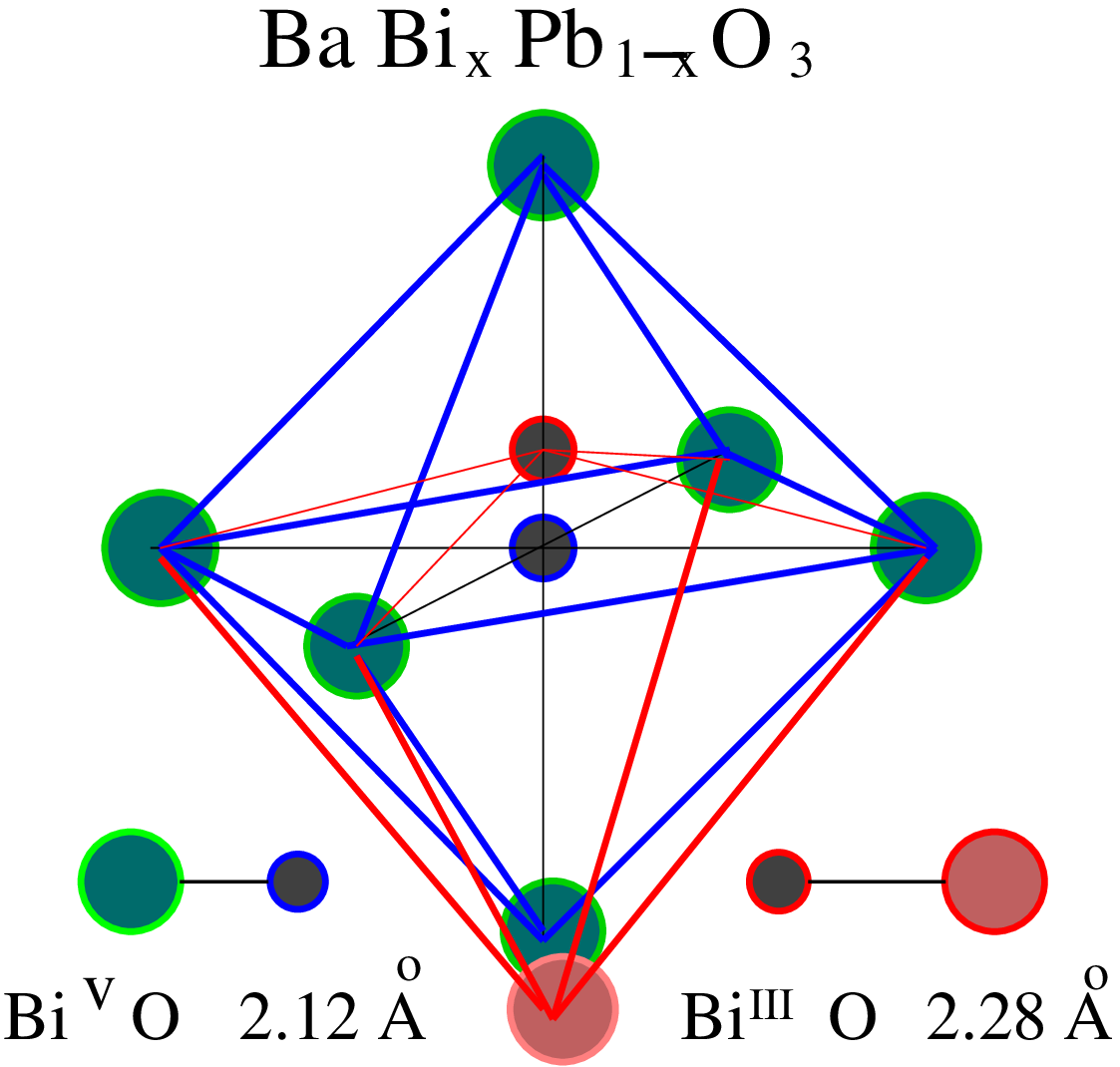}
\caption{The two stable Bi$^{V}$ and Bi$^{III}$ configurations in regular and pseudo octahedral 
ligand environments (after Ref.~\protect\refcite{Simon-1988}).}
\label{Fig2}
\end{minipage}
\end{figure}

The long standing  efforts to synthesize  superconductors with critical temperatures higher than about 
25~K gradually  faded away as the decades past. Three quarters of a century after the first discovery 
of superconductivity by Kamerlingh-Onnes\cite{Kamerlingh-Onnes-1911} in 1911 (Hg with a $T_c$ 4.25 K), 
a ceramic material of not particularly good materials qualities surfaced in 1986 and Bednorz and 
Mueller produced this long searched after goal.\cite{Bednorz-Mueller-1986} The inhomogeneities of 
these materials, which seem to be intrinsic, are  related to local dynamical lattice instabilities, 
and could be a prime factor to  bypass the limitation of $T_c$ \cite{Anderson-Matthias-1964} in  
classical low temperature BCS type phonon-mediated superconductors.\cite{BCS-1957} Local dynamical 
lattice instabilities are known to trigger diamagnetic fluctuations\cite{Vandenberg-Matthias-1977} 
without leading to any global translational symmetry breaking, which would kill the superconducting 
state and end up in a charge ordered phase. Maximal values of $T_c$  can be expected for materials, 
which are synthesized  at the highest temperature in the miscibility phase diagram,\cite{Sleight-1991} 
see Fig.~\ref{Fig1}. They present thermodynamically stable single-phase solutions at the boundary of insulating 
and metallic regions, which upon cooling condense into single-phase solid solutions. Their high 
kinetic stability prevents them from decomposing  into the different compositions of the mixture one 
started with, due to a freezing-in of the high entropy mismatch of thermodynamically stable phases 
in the synthesization process. Microscopically, their metastability arises
from adjacent cation-ligand  complexes with incompatible inter-atomic distances.

Following such a strategy in material preparations, BaBi$_x$Pb$_{1-x}$O$_3$,\cite{Sleight-1975} Ba$_{1-x}$K$_x$BiO$_3$ \cite{Cava-1988,Hinks-1988} and Pb$_{1-x}$Tl$_x$Te,\cite{Chernik-1981,Moizhes-1983} just to mention a few representative examples, were synthesized and exhibited  superconducting phases with 
$T_c$~=  13, 30, 1.5~K. These are relatively high values, considering their small carrier density of around 10$^{-20}$/cm. Their  parent compounds are respectively BaBiO$_3$, which is a diamagnetic insulating charge ordered state and PbTe, which is a  small gap semiconductor.  Their non-metallic phases are turned into superconductors upon  partially substituting Bi by Pb, Ba by K and 
Te by Tl. According to band  theory,  BaBiO$_3$  should be metallic, composed of exclusively Bi$^{4+}$. But the high polarizibility of the oxygens\cite{Hase-2007} makes it synthesize in a mixture of  Bi$^{5+}$ and Bi$^{3+}$, rendering Bi$^{4+}$ an unstable valence state. Such materials are sometimes referred to as ``valence skippers''. Bi$^{5+}$ occurs in a regular octahedral ligand environment, with a Bi--O distance of 2.12~\AA. Bi$^{3+}$, on the contrary, occurs  in a pseudo-octahedral ligand environment, with one of the oxygen ions in the octahedral being displaced to such an extent (with a corresponding Bi--O distance of 2.28~\AA) that it effectively becomes O$^{2-}$, after having transferred an electron to Bi$^{4+}$.\cite{Simon-1988} This charge disproportionation results in negative U centers - the Bi$^{3+}$ sites on which electrons pair up. The undoped parent compound  stabilizes in a translational symmetry broken state of alternating Bi$^{3+}$ and Bi$^{5+}$ ions. Upon doping (partially replacing Bi by Pb), the system becomes superconducting with locally fluctuating [Bi$^{3+}\Leftrightarrow$Bi$^{5+}$], see Fig.~\ref{Fig2}. The situation is similar in Pb$_{1-x}$Tl$_x$Te. Its parent compound, PbTe, involves  unstable Te$^{2+}$ valence cations, which disproportionate into Te$^{1+}$ and Te$^{3+}$ ions, with the first again acting as negative U centers. Upon doping PbTe $\rightarrow$ Pb$_{1-x}$Tl$_x$Te it changes into a superconductor, driven by those negative U centers.

\section{The scenario}

\begin{figure}[tbp]
\begin{center}
\includegraphics[width=8.25cm]{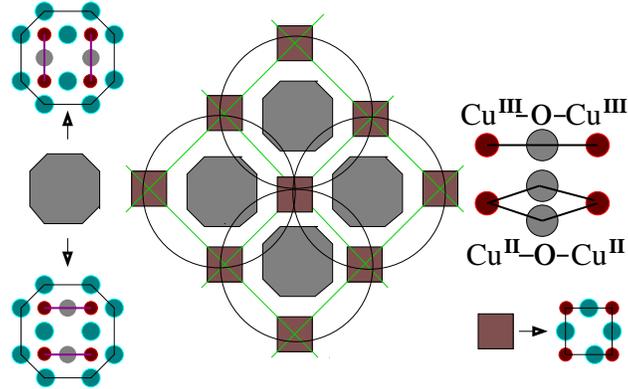}
\end{center}
\caption{Schematic view of the local segregation of the CuO$_2$ planes 
into rotational symmetry breaking nano-size clusters (grey) (after Kohsaka \textit{et al.} Ref.~\protect\refcite{Kohsaka-2008}), 
containing dynamically deformable Cu-O-Cu bonds which act 
as pairing centers. These pairing centers  are  embedded in a metallic matrix of Cu$_4$O$_4$ 
square (brown) plaquettes, on which the charge carriers circle around the central trapping centers, 
indicated by black circles. The small red filled circles indicate the Cu ions and the blue filled 
larger ones the oxygens. The fluctuating bridging oxygens are indicated by grey filled circles.}
\label{Fig3}
\end{figure}

Cuprates High $T_c$ materials bare some similarity to such socalled valence skipping compounds.
Upon doping, their anti-ferromagnetic half-filled band Mott-insulating parent compound, composed 
of Cu$^{II}$ - O  - Cu$^{II}$ bonds in the CuO$_2$ planes, becomes dynamically locally  unstable 
and leads to  dynamically fluctuating  
Cu$^{II}$ - O  - Cu$^{II} \Leftrightarrow$ Cu$^{III}$ - O  - Cu$^{III}$ 
bonds.\cite{Kohsaka-2007,Gomes-2007} Different from the socalled valence
skippers mentioned above, the basic building stones are covalent molecular 
bonds. This does  not imply Cu valencies of 
2+, respectively 3+. The notation II and III indicates stereochemical 
configurations defined by intrinsic Cu-O bond-lengths, which are 
1.93~\AA for Cu$^{II}$ - O  - Cu$^{II}$  and 1.83~\AA for
Cu$^{III}$ - O  - Cu$^{III}$. Because of stereochemical misfits between the 
CuO$_2$ planes and the surrounding cation-ligand complexes in the 
neighboring insulating planes, acting as charge reservoirs, the Cu - Cu 
distance of the Cu$^{II}$ - O  - Cu$^{II}$ bonds  in a square 
planar oxygen ligand environment are  forced to reduce their lengths. They do  that by 
a bond buckling, where  the bridging oxygen of those bonds moves out of the CDuO$_2$ plane and 
thus respects its attributed stereochemical distance. Upon hole doping, the static bond buckling 
becomes dynamic, involving unbuckled  linearly Cu$^{III}$-O-Cu$^{III}$ 
bonds. The bridging oxygens then fluctuate in and out of the CuO$_2$  plane 
and thus dynamically modulate the length of the Cu-O bonds. Experimentally, this 
feature is manifest  in a splitting of the corresponding local  bond stretch 
mode, which sets in in low doped metallic regime and disappears upon 
overdoping.\cite{Reznik-2006} The two stereo-chemical configurations 
Cu$^{II}$ - O  - Cu$^{II}$ and Cu$^{III}$ - O  - Cu$^{III}$, observed in 
EXAFS,\cite{Zhang-2009} differ by two charges and it is that which results in the 
dynamically fluctuating diamagnetic bonds, which are the salient features of phase 
fluctuation driven superfluidity of the cuprates.\cite{Uemura-1989} Inspite of  
the huge differences in the Cu-O bond-length $\simeq 0.1$~\AA of the Cu-O-Cu constituents, 
the cuprates do not break the overall translational symmetry caused by a charge order, but prefer to 
segregate into a local distribution of such bonds. They stabilize in a checkerboard 
structure of nano-size clusters, surrounded by small molecular Cu$_4$O$_4$ sites, such that on
a local basis those systems form a bipartite structure (see an idealized form for it in Fig.~\ref{Fig3}).
The dynamically fluctuating uni-directional Cu - O - Cu bonds exist in two equivalent orthogonal 
directions and thus break not only the local translational but also rotational symmetry.\cite{Kohsaka-2008}
This prevents any tendency to  long range translational symmetry breaking.

One might  have expected such a result from the outset. Superconductivity in 
the cuprates is destroyed by phase fluctuations of the order parameter, without 
that its amplitude would suffer any significant depreciation.\cite{Emery-Kivelson-1995,Franz-1998} 
This can only happen when Cooper pairs are fairly local entities, such as to provide 
superfluidity like  features with a $T_c$ scaling with the 
superfluid density at zero temperature\cite{Uemura-1989} and  XY 
characteristics of the transition,\cite{Salamon-1993} in strong contrast
to  the mean field characteristics of standard amplitude fluctuation controlled 
BCS superconductivity. The local nature of the Cooperons also implies that 
pairing correlations must develop already well above $T_c$. Because of the charged 
and local nature of the particles involved in this pairing, its has a strong effect on 
the local lattice deformations to which they are irrevocably coupled. 

In principle one can now envisage two scenarios: \\
(i) if the coupling is very strong, the charge carriers form bipolarons, 
which in principle can condense into a superfluid phase - the 
Bipolaronic Superfluidity,\cite{Alexandrov-1981} but which are more likely to 
end up in an insulating state of statically disordered localized
bipolarons, if not a charge ordered state of them.\cite{Chakraverty-1998} \\
(ii) if the the coupling is not quite as strong, the charge carriers will be 
in a mixture of quasi-free itinerant states and localized states, where they 
are momentarily selftrapped in form of local bound pairs. Local dynamical pairing then 
occurs at some temperature $T^*$, at which the single-particle
density of states develops a pseudogap, because of the electrons getting paired up on a finite time 
scale.\cite{Ranninger-1995} Upon lowering the temperature, those phase incoherent locally 
dynamically fluctuating  pairs acquire short range phase coherence, which ultimately leads to 
their condensation and phase locking in a superfluid state at some temperature $T_{\phi}$, 
exclusively controlled by phase fluctuations.\cite{Cuoco-2004} Experimental evidence for that is 
now well established in terms of the transient Meissner effect,\cite{Corson-1999} 
the Nernst effect in thermal transport,\cite{Xu-2000} the torque measurements 
of diamagnetism\cite{Wang-2005} and a proximity induced pseudogap of metallic 
films deposited on La$_{2-x}$Sr$_x$CuO$_4$.\cite{Yuli-2009}

Standard low temperature BCS superconductors are controlled by amplitude fluctuations,
 with the feature that at $T_c$ the expectation value of the amplitude vanishes, i.e.,  
 the number of Cooper pairs goes to zero and thus the phase of the order parameter becomes 
 redundant. For superfluid He, the opposite is the case. The onset of superfluidity is 
 controlled exclusively by spatial phase locking of the bosonic order parameter.
 The amplitude there is fixed, since the number of bosonic He atoms is conserved. The 
 superfluid state is destroyed at a certain $T_{\phi}$, given by the phase stiffness
 of the condensate. The high $T_c$ superconductors lie between those two limiting cases. 
 Amplitude and phase fluctuations are then strongly inter-related and  have to be treated
 on equal footing. The onset of a finite amplitude of the order parameter happens at a 
 temperature $T^*$, while that of the phase locking of the bosonic Cooperons, signaling
 the onset of superconductivity,  occurs at $T_{\phi}$, which can be well below $T^*$.  Dealing with 
 such a situation requires to disentangle the description of phase and amplitude fluctuations of the
 Cooperons. Cooperons are neither true bosons nor hard-core bosons, but their center of mass  
 can be associated to bosons in real space. Using a renormalization procedure one can project 
 out such bosons of the Cooperons, leaving us with remnants arising from kinematical 
 interactions  due to their non-bosonic statistics coming from their internal fermionic degrees of 
 freedom. That leads to a damping of such bosons. Those bosons have an amplitude and
 a phase and hence can account for both:  (i)~their dissociation into fermion pairs at $T^*$,
 where their amplitude disappears together with the closing  of the pseudogap in the 
 fermionic single-particle subsector and (ii)~the breaking up their spatial phase coherence
 at $T_{\phi}$, where the bosons enter into a phase  disordered  boson metallic state, albeit
 keeping short range phase coherence  in a finite temperature regime above $T_{\phi}$. This 
 scenario follows some old ideas in the early days of the BCS theory, when exploring the  
 interaction between the electrons and the Cooper pairs in a perturbative way,\cite{Kadanoff-1961}
 or, in a more recent study, using a phenomenological  picture for strong inter-relations between the 
 two.\cite{Tchernyshyov-1997} In the approach we have been following now for several years, 
 the inter-relation between phase and amplitude fluctuations is cast into Andreev scattering 
 of the fermions on  bosons. It results in a partial trapping of the 
 fermions in form of localized  bound fermion pairs in real space, and leads to fermionic
 excitations, which are in a superposition of itinerant and localized states, consistent  
 with recent STM imaging results.\cite{Kohsaka-2008} The pairing we are considering here
 corresponds to a resonant pairing, induced by fluctuating molecular bonds driven by intrinsic 
 dynamical lattice instabilities. The mechanism is similar to that of a Feshbach 
 resonance\cite{Feshbach-1958} in fermionic atom gases, where fermionic atoms exist simultaneously 
 in form of scattering electron-spin singlet states and as bound electron-spin triplet states.

\begin{figure}[tbp]
\begin{center}
\includegraphics[width=8cm]{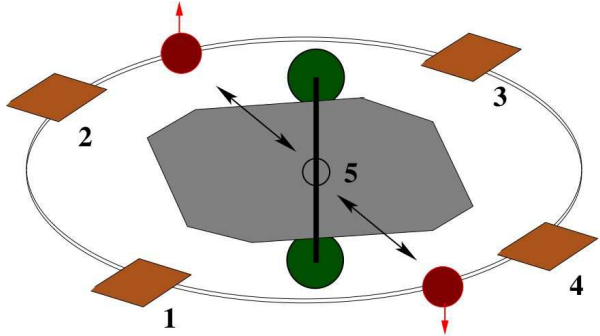}
\end{center}
\caption{Simplified version of a local cluster of Fig.~\ref{Fig3}, composed of (i)~a trapping domain 
Cu$_4$O$_{10}$, replaced here by a deformable dumbbell oscillator (green filled circles) at 
site 5, capable of trapping two electrons (red filled circles) in form of a bipolaron and (ii)~four  Cu$_4$O$_4$ square plaquettes, which are replaced by effective sites 1 - 4, on which the electrons circle around the central site 5.}
\label{Fig4}
\end{figure}

\section{The model}

Resonant pairing systems can most efficiently be described by  a 
Boson Fermion Model (BFM), such as I had conjectured  shortly after 
the proposition of the Bipolaronic Superconductivity.\cite{Alexandrov-1981} 
This  phenomenological model represents a mixture of itinerant fermions and localized
bosons ---locally bound fermion pairs--- with an exchange interaction between 
the two. A common chemical potential is assumed such as to assure that we are 
dealing with a unique species of charge carriers. They must exist simultaneously 
in (i)~itinerant states circling around the trapping centers and (ii)~as localized states
on the  trapping centers, the moment they  pair up. This picture is quite general and does not 
depend on any specific mechanism for pairing. It can arise from holes in a strongly correlated  
Hubbard system close to half filling,\cite{Anderson-1987} where 
the exchange happens between hole singlet-pairs and itinerant uncorrelated
holes\cite{Altman-2002} or, as we have been pursuing it for years, 
from a polaronic mechanism. 

Our picture of hole-doped cuprates is based on an ensemble of effective sites, embedded in a 
crystalline structure, each one being composed of a deformable cation-ligand complex, surrounded by an 
undeformable local environment, the four-site rings, see Fig.~\ref{Fig3}.  The local environments of 
neighboring  deformable molecular clusters overlap with each other, which assures a potentially 
metallic phase of the charge carriers. If the coupling of electrons into bosonic bound pairs 
outweighs their kinetic energy, it leads to a Bose liquid, which can be either superfluid or 
insulating, without that any fermions are present in the background. High $T_c$ materials do however 
exhibit such a background Fermi sea and at the same time a transition into a superfuid state which is
governed by the dynamics of bosonic fermion-pairs rather than by the opening up of a gap in the 
electronic single-particle density of states. Itinerant bosons are induced upon hole 
doping. They  form and disappear spontaneously on the deformable cation-ligand  clusters and 
thereby break the crystal symmetry on a local level. This resonating pairing originates
from the local dynamical lattice instabilities and metastability of the cuprates. It implies 
that those systems are intrinsically inhomogeneous, having segregated, upon doping, into effective 
sites of molecular clusters,  partially occupied  by such bosons.  Resonating pairs of 
fermions, implies electrons fluctuating in and out of those deformable clusters, on which they 
self-trap themselves in form of bound pairs
on a finite time scale. This situation is energetically more favorable than that of purely 
localized bosonic bound fermion pairs or itinerant uncorrelated fermions. It benefits 
from a lowering of the ionic level due to a polaron level shift and from the 
bound fermion pairs acquiring itinerancy. 

Let us now demonstrate that on the basis of a small cluster calculations, following a previous
detailed study\cite{Ranninger-2008} on that. The segregation of the CuO$_2$ plane charge distribution, 
upon hole doping,  has been seen in STM imaging studies\cite{Kohsaka-2008} and a sketch for such 
segregation is illustrated in Fig.~\ref{Fig3}. It consists of nano-size Cu$_4$O$_{10}$ domains composed of 
deformable Cu - O - Cu bonds which are capable of trapping momentarily two itinerant holes 
from the surrounding Cu$_4$O$_4$ plaquettes. Those latter form the remnant back bone 
structure corresponding to the undoped cuprates carrying the itinerant charge carriers. 
The fluctuation of the charge carriers back and forth between the two subcomponents of 
the segregated CuO$_2$ planes ultimately  induces a diamagnetic component among the itinerant 
charge carriers. 

\section{Resonant pairing and local dynamical lattice instabilities}

\begin{figure}[h!t]
\begin{minipage}[t]{6cm}
\includegraphics[width=6cm]{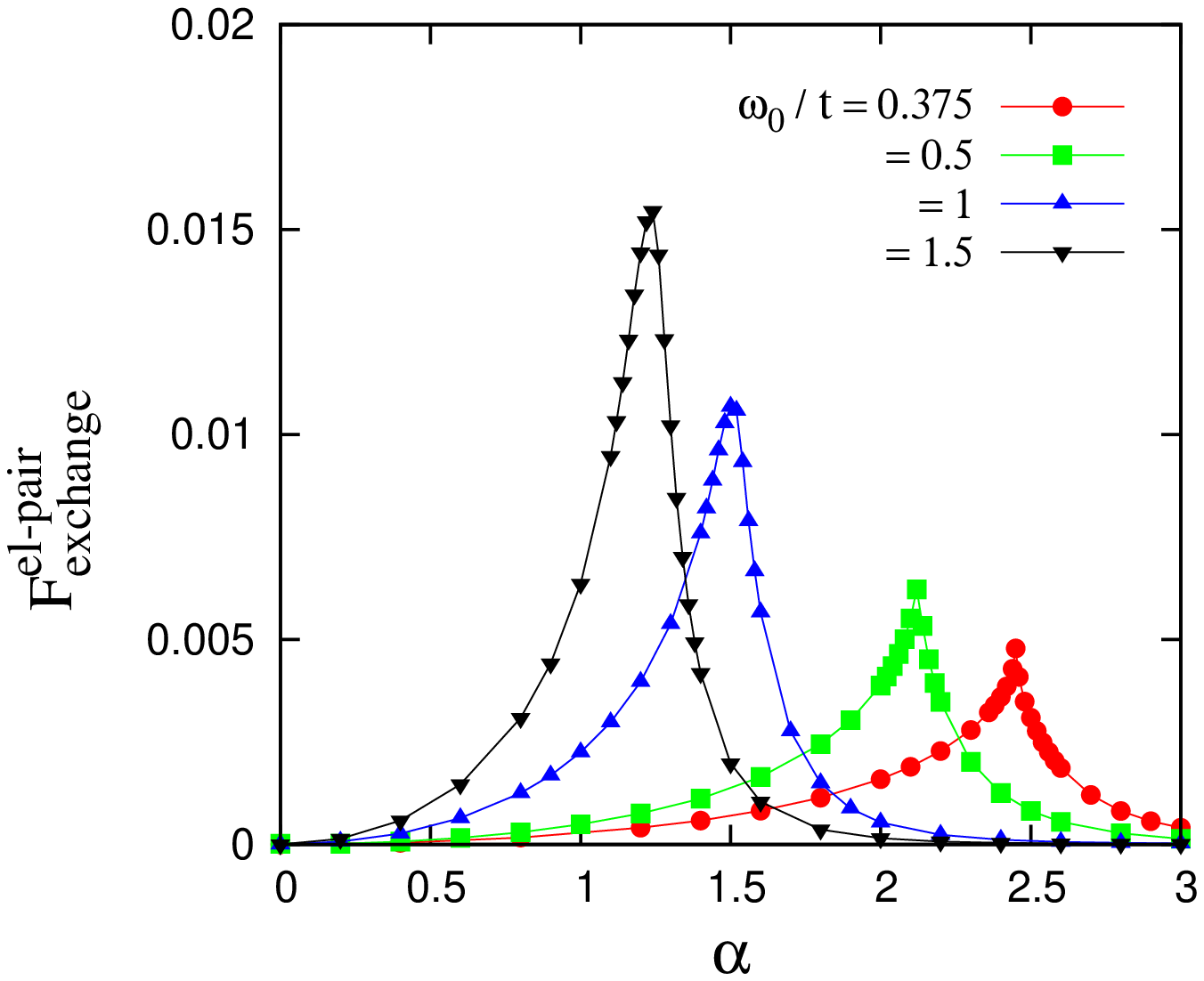}
\caption{(Color online) The efficiency factor $F^{\rm pair}_{\rm exch}$ as a function 
of $\alpha$ for converting a localized bipolaron on the central polaronic  complex into a pair of diamagnetically correlated electrons on the ring, for several adiabaticity ratios $\omega_0/t$ (after Ref.~\protect\refcite{Ranninger-2008}).}\label{Fig5}
\end{minipage}
\hfill
\begin{minipage}[t]{6cm}
\includegraphics[width=6cm]{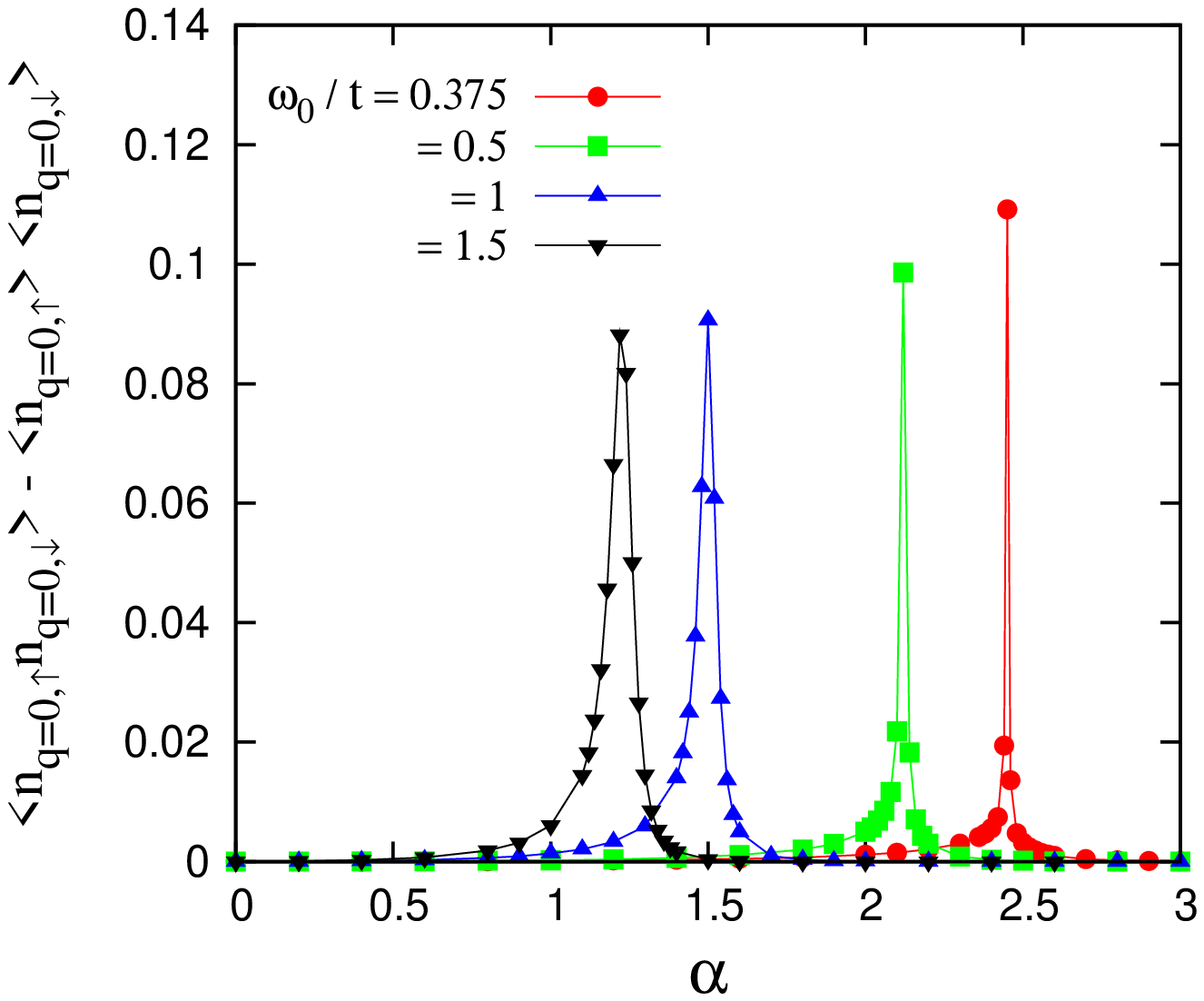}
\caption{(Color online) Density $n_p$ of diamagneticaly correlated electron pairs on the ring as a function of $\alpha$ showing a strong enhancements for $\alpha = \alpha_c$ for a set of adiabaticity ratios $\omega_0/t$ (after Ref.~\protect\refcite{Ranninger-2008}).}\label{Fig6}
\end{minipage}
\end{figure}

The physics elaborated above is an intrinsic property of the nano-size  Cu$_4$O$_{10}$ clusters, 
making up those materials. We illustrate that by mapping the central Cu$_4$O$_{10}$ trapping domain 
together with its four surrounding Cu$_4$O$_4$ plaquettes, into a tractable model: a central effective 
site given by a deformable oscillator surrounded by four atomic sites (see Fig.~\ref{Fig4}). On the first, 
the itinerant charge carriers can get self trapped into localized 
bipolarons, corresponding to a deformation which mimics the process 
Cu$^{III}$ - O  - Cu$^{III} \Leftrightarrow$ Cu$^{II}$ - O  - Cu$^{II}$ in the 
trapping domains of Fig.~\ref{Fig3}. Resonant pairing occurs when the energies 
of the bipolaronic level corresponds to the Fermi energy of the itinerant 
fermionic subsystem. In our toy model we shall characterize these states 
by two-electron eigenstates on the four-site ring.  
The Hamiltonian for such a small cluster system is given by
\begin{eqnarray}
H & = & - \, t \sum_{i \ne j = 1 ... 4, \,\sigma} \left[
c^{\dagger}_{i,\sigma} c_{j,\sigma}
\, + \, h.c.  \right]  
 -\, t^* \sum_{i=1 ... 4, \,\sigma} \left[ c^{\dagger}_{i,\sigma}
c_{5,\sigma} \, + \, h.c.  \right] \, \nonumber \\ 
& + & \, \Delta \, \sum_{\sigma} c^{\dagger}_{5,\sigma} c_{5,\sigma}  
 + \, \hbar\omega_0 \left[a^{\dagger} a + \frac{1}{2}\right] \,
- \hbar\omega_0\alpha \, \sum_{\sigma} \, c^{\dagger}_{5,\sigma}
c_{5,\sigma}\, \left[a + a^{\dagger}\right],
\end{eqnarray}
where $c^{(\dagger)}_{i\sigma}$ denotes the annihilation (creation)
operator for an electron with spin $\sigma$ on site $i$, and
$a^{(\dagger)}_5$ the phonon annihilation (creation) operator
associated with a deformable cation-ligand complex at site $5$.
$\alpha$ denotes the dimensionless electron-phonon coupling constant, 
$\omega_0$ the Einstein oscillator frequency of the local dynamical 
deformation on the central polaronic site 5, $t$ the intra-ring hopping 
integral and $t'$ the one controlling the transfer between the ring and the central
site 5. $\Delta$ denotes the bare energy of the electrons when sitting on site 
5. before they couple to the local lattice deformation.
This local cluster describes the competition between localized
bipolaronic electron pairs on the central cation-ligand complex
and itinerant electrons on the plaquette sites, when the energies
of the two configurations are comparable with each other, i.e. for $2\Delta - 4
\hbar \omega_0 \alpha^2  \simeq -4t$. There is a narrow resonance regime
where this happens, as we can see from the strong enhancement of the 
efficiency rate $F^{\rm pair}_{\rm exch}$ 
\begin{eqnarray}
F^{\rm pair}_{\rm exch} = \langle
GS|c^{\dagger}_{5,\uparrow}c^{\dagger}_{5,\downarrow}
c_{q=0,\downarrow}c_{q=0,\uparrow}| GS \rangle - \langle GS|c^{\dagger}_{q=0,\uparrow}
c_{5,\uparrow}|GS \rangle\langle GS|c^{\dagger}_{q=0,\downarrow}c_{5,\downarrow}|GS \rangle
\end{eqnarray}
for transferring electron pairs back and forth between the ring sites and the central polaronic
site,\cite{Vandenberg-Matthias-1977} (see Fig.~\ref{Fig5}) and where we have subtracted out the exchange due to incoherent single-electron
processes. $|GS\rangle$ denotes the ground state of the cluster. This resonant scattering process 
induces a diamagnetic component of the electrons moving on the ring. The density $n_p$ of such 
diamagnetically correlated resonantly induced pairs is given by 
\begin{equation}
n_p = \langle n_{q=0,\uparrow}n_{q=0,\downarrow}\rangle - \langle
n_{q=0,\uparrow}\rangle \langle n_{q=0,\downarrow}\rangle
\end{equation}

The variation of $n_p$ with $\alpha$ (Fig.~\ref{Fig6}) shows an equally sharp enhancement 
near the resonance $\alpha_c$, which depends on the adiabaticity ratio $\omega_0 / t$. 
Its relative density is quite sizable - about $20\%$ of the total average density
of electrons on the ring at $\alpha_c$. Slightly away from
this resonant regime, the electrons are either pair-uncorrelated
(for $\alpha \leq \alpha_c$) or not existent on the plaquette (for
$\alpha \geq \alpha_c$), because of being confined to the polaronic
cation-ligand complex as localized bipolarons.

\begin{figure}[tbp]
\begin{center}
\includegraphics[width=0.8\textwidth]{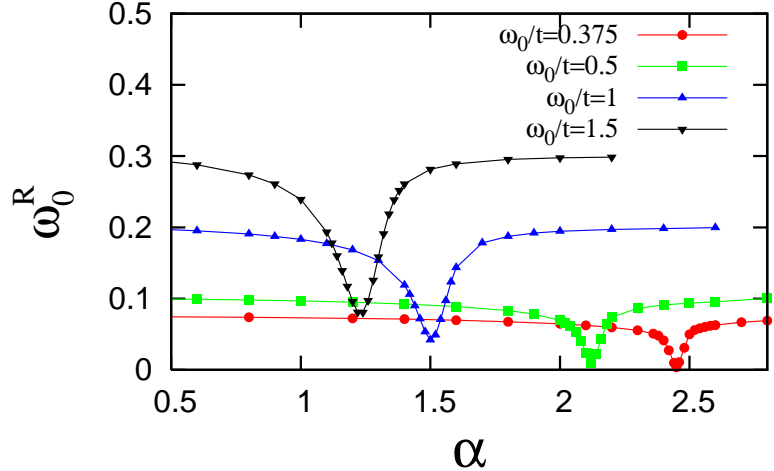}
\end{center}
\caption{(Color online) The phonon softening  as a function of the electron-dimer coupling 
constant for various adiabaticity ratios $\omega_0/t$~= 0.375, 0.5, 1 and 1.5 (after Ref.~\protect\refcite{Ranninger-2008}).}
\label{Fig7}
\end{figure}

The features driving these electronic exchange are local deformations of the molecular 
dimer, which induce  a correlated dynamics of the local dimer deformations  
and the local charge  fluctuations. Denoting the modulation in length of the dimer by 
$X = \sqrt{\hbar/2M\omega_0}[a^+ + a]$, the corresponding phonon Greens function is
\begin{equation}
D_{ph} = {2M \omega_0 \over \hbar} \langle\langle X;X \rangle\rangle,
\end{equation}
where $M$ denotes the mass of the vibrating dimer atoms (green in Fig.~\ref{Fig4}).
Our previous results\cite{Ranninger-2008} on this local lattice dynamics are reproduced 
in Fig.~\ref{Fig7}, where we illustrate  the softening of the bare phonon mode 
$\omega_0$ down to $\omega_0^R$ at the resonance. It is this frequency which determines the 
correlated charge-deformation fluctuations and provides us with an estimate for the effective 
mass of the diamagnetic pairs (see the discussion in the SUMMARY section of Ref.~\refcite{Ranninger-2008}). 

\section{Separation of phase and amplitude variables}

Molecular clusters, like the one discussed in  the previous section, form the building blocks of hole
doped segregated CuO$_2$ planes and which, accordingly assembled, form the macroscopic 
cuprate structure: a regular bipartite lattice structure, idealized in Fig.~\ref{Fig3}. This picture is
in accordance with the checker-board structure seen in  STM imaging studies.\cite{Kohsaka-2008}
The local stability of such dynamically fluctuating molecular clusters is obtained at resonance, 
i.e., at $\alpha_c$, which depends on the adiabaticity ratio but more sensibly on the exchange rate 
$\propto\omega_0^R$ between the localized pairs and the itinerant electrons. This latter will 
strongly depend on the degree of hole doping.

The microscopic mechanism for hole doping in the cuprates is a complex phenomenon and far 
from being understood. The undoped systems present a homogeneous lattice structure with buckled 
Cu$^{II}$ - O - Cu$^{II}$ molecular bonds. Due to strong Hubbard correlations among the electrons 
in such a half-filled band system, the ground state is an antiferromagnetic insulator. Upon slight 
hole doping it breaks down into a glassy heterogeneous structure, which we interpret in our 
scenario as a phase uncorrelated bipolaronic Mott insulator,\cite{Cuoco-2004,Stauber-2007} composed of 
Cu$^{II}$ - O - Cu$^{II}$ covalent molecular bonds, behaving as hardcore intersite bipolarons. Upon 
increasing the hole doping beyond around 0.10 per Cu ion, the system segregates in to a checkerboard 
structure with nano-size molecular Cu - O - Cu clusters, in a mixture of Cu$^{II}$ - O - Cu$^{II}$  
and Cu$^{III}$ - O - Cu$^{III}$ molecular bonds. These nano-size clusters act as trapping centers 
for the itinerant electrons moving on the sublattice in which those cluster are embedded (see Fig.~\ref{Fig3}).
Given the experimental findings on the effect of hole doping on the crystal structure monitored by an 
increase in covalency,\cite{Roehler-2004-2009} 
we conjecture that hole doping primarily induces hole-bipolarons (Cu$^{III}$ - O - Cu$^{III}$ molecular 
bonds) on the nano-size trapping centers. This triggers local charge fluctuations, by momentarily 
capturing two electrons from of the surrounding four-site ring environment, transforming 
Cu$^{III}$ - O - Cu$^{III}$ into Cu$^{II}$ - O - Cu$^{II}$ molecular bonds. The  two initially 
uncorrelated electrons thereby acquire a diamagnetic component, which they keep upon returning into 
the itinerant subsystem  via the inverse process 
[Cu$^{II}$ - O - Cu$^{II}]$ $\rightarrow$ [Cu$^{III}$ - O - Cu$^{III}$].

This resonant back and forth scattering between bound and unbound electrons stabilizes 
the dynamical transfer of a small fraction of electrons $n_F^h$ from the initially quasi half-filled band 
of itinerant charge carriers into the trapping centers.
As a result, the Fermi surface shrinks, which implies a hole density $n_F^h = n_B$ of 
single-particle excitations. Since  $n_F^h$ varies only in a small regime, between 0.10 and 
0,22, covering the superconducting phase, we choose for our resonant pairing scenario, a total number of 
charge carriers around $n_{\rm tot} = n_F + n_B \simeq 1$, which  will slightly decrease with hole doping. 
This will lead to a self-regulating boson density, such as $n_B \leq 0.08$ for the optimally
doped case $n_F^h \simeq 0.16$. The electrons, being momentarily trapped into bound electron pairs 
on the deformable molecular clusters, can be  be associated to bosonic variables $b^{(\dagger)}$ with
$n_B$~= $\langle b^{\dagger}b\rangle$.
Our further analysis is based on the BFM which, in momentum space, is given by the Hamiltonian

\begin{eqnarray}
H_{\rm BFM} &=& H^0_{\rm BFM} + H^{\rm exch}_{\rm BFM}\label{eq5} \\
H^0_{\rm BFM} &=& \sum_{{\bf k}\sigma}(\varepsilon_{\bf k} -\mu)c_{{\bf k}\sigma}^{\dagger} 
c_{{\bf k}\sigma}+\sum_{\bf q} 
(E_{\bf q}-2\mu) b_{\bf q}^{\dagger} b_{\bf q}.\qquad\label{eq6}\\
H^{\rm exch}_{BFM}&=&(1/\sqrt{N})\sum_{{\bf k},{\bf q}}(g_{{\bf k},{\bf q}}b_{\bf q}\dag 
c_{{\bf q-k},\downarrow}c_{{\bf k},\uparrow}+H.c.) \label{eq7}\\
H^{F-F}_{\rm BFM} &=& {1  \over N} \sum_{{\bf p},{\bf k},{\bf q}} U_{{\bf p},{\bf k},{\bf q}}
c^{\dagger}_{{\bf p} \uparrow}c^{\dagger}_{{\bf k} \downarrow}c_{{\bf q} \downarrow}
c_{{\bf p}+{\bf k}-{\bf q} \uparrow}.
\label{H_BFM}
\end{eqnarray}

Following the basic electronic structure of the cuprates, we assume a d-wave symmetry for the 
boson-fermion exchange interaction  $g_{{\bf k},{\bf q}}$~= $g[\cos k_x  - \cos k_y]$, between (i)~pairs 
of itinerant charge carriers $c^{(\dagger)}_{{\bf k}\sigma}$ circling around the polaronic sites and (ii)~
the polaronicaly bound pairs of them $b_{\bf q}^{(\dagger)}$, when these same charge carriers have hopped 
onto this site and got self trapped. The anisotropy of the itinerant charge carrier dispersion 
is assured  by the standard expression $\varepsilon_{\bf k}$~= $-2t[\cos k_x + \cos k_y] + 4t'\cos k_x \cos k_y$ 
for the CuO$_2$ planes with $t'/t$~= 0.4. The strength of the boson-fermion exchange coupling in our 
effective Hamiltonian can be estimated  from our study of the single cluster
problem and is given by $F^{\rm pair}_{\rm exch}$, which is  related to the electron-lattice
coupling $\alpha$  and the bare phonon frequency $\omega_0$, as discussed in detail in Ref.~\refcite{Ranninger-2008}.
In order to get an insight into the inter-related amplitude and phase fluctuations in hole doped High 
$T_c$ cuprates, we transform this Hamiltonian by a succession of unitary transformations into a block 
diagonal form, composed of  exclusively (i)~renormalized fermionic particles and (ii)~renormalized
bosonic particles. This gives us an access to study the characteristic single-particle
spectral properties  such as the pseudogap  and the two-particle-properties, controlling
the onset of phase locking and diamagnetic fluctuations, which govern the transport as 
one approaches $T_c$ from above. Using Wegner's Flow Equation renormalization
procedure\cite{Wegner-1994} we eliminate the boson-fermion exchange coupling in
successive steps, obtaining a Hamiltonian of the same structure as the initial 
one, Eqs~(\ref{eq5}--\ref{H_BFM}), but with renormalized parameters  
$g^*=0$, $\varepsilon^*_{\bf k}$, $E^*_{\bf q}$, $U^*_{{\bf p},{\bf k},{\bf q}}$, 
which have evolved out of the bare ones we started with, i.e., 
$g, \varepsilon_{\bf k}, E_{\bf q}$~= $\Delta-\mu, U_{{\bf p},{\bf k},{\bf q}} = 0$. 
The chemical potential  $\mu$, which also evolves in the course of this renormalization 
procedure is determined at each step of it, such as to assure a given total density 
of fermionic and bosonic particles $n_{\rm tot}$. From our detailed calculations\cite{Ranninger-2009}
of this renormalized Hamiltonian and its spectral properties we see a transformation of
the electron dispersion, close to the chemical potential,  which changes into an S-like shape
upon lowering the temperature below a certain $T^*$. It signals the 
onset of pairing and  manifests itself in a corresponding opening of the pseudogap 
in the single-particle density of states. Simultaneously the intrinsically dispersion-less 
bosons acquire a $q^2$ like spectrum, albeit overdamped until, upon further decreasing the
temperature, they finish up as well defined quasi-particle which ultimately condense into a 
superfluid phase below $T=T_{\phi}$. Their  $q^2$ like spectrum transforms 
into a linear in $q$ branch, which signals the phase locked superfluid state of those bosons.
 
\begin{figure}[tbp]
\begin{center}
\includegraphics[width=9.15cm]{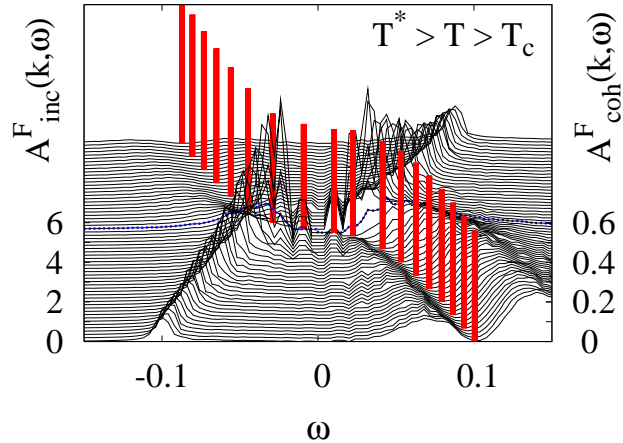}
\end{center}
\caption{The single-particle spectral function $A({\bf k},\omega)$~=
 $A_{\rm inc}({\bf k},\omega) + A_{\rm coh}({\bf k},\omega)$ in the 
pseudogap phase around the hidden ``Fermi surface'' ($\omega=0$) and near the anti-nodal point. 
The set of curves corresponds to different ${\bf k}$ vectors, orthogonally intersecting the hidden
Fermi surface. The curve in blue indicates the spectrum exactly at this hidden Fermi surface 
(after Ref.~\protect\refcite{Ranninger-2009}).}
\label{Fig8}
\end{figure}

The anisotropic d-wave structure of the boson-fermion exchange coupling implies a variation
of its strength, going from  zero at the nodal points $[\pm \pi/2,\pm \pi/2]$ of the Fermi surface
toward its maximal value, equal to g, near the hotspots ${\bf k}$~= $[0, \pm \pi]$, $[\pm \pi,0]$. Upon 
moving on an arc in the Brillouin zone, corresponding to the chemical potential, i.e.,
$\varepsilon^*(k_x,k_y)$~= $\mu$, one observes a well defined Fermi surface around the nodal
points. But upon parting from this limited region, the Fermi surface gets 
hidden because of the onset of a pseudogap. This has been called "Fermi Surface Fractionation". 
It implies in this region of the Brillouin zone (see Fig.~\ref{Fig8}) three astonishingly related features:
(i)~diffusively 
propagating Bogoliubov branches below $\omega = 0$ for ${\bf k} \geq {\bf k}_F$, indicated by 
$A^F_{inc}({\bf k},\omega)$ (black lines). They  are remnants of superconducting phase locking on a 
finite space-time scale, which have been predicted by us in 2003\cite{Domanski-2003} and only very 
recently have been verified  by ARPES in 2008,\cite{Kanigel-2008}; (ii)~localized high energy modes 
above $\omega=0$ (black) for ${\bf k} \geq {\bf k}_F$, which occur together with those diffusively 
propagating Bogoliubov modes.  We have interpreted them as the internal degrees freedom 
of those two-particle collective phase modes, showing up in  their high frequency response. They 
present locally partially trapped single-particle states making up the bipolaronic 
Cooperons,\cite{Ranninger-2009} which include; (iii)~in-gap single-particle contributions, denoted by 
$A^F_{\rm coh}({\bf k},\omega)$ (red delta like peaks). Their weight should be understood as the overall
spectral weight for very broadened peaks, as we know from independent studies, such as 
self-consistent perturbative treatments\cite{Ranninger-1995} and DMFT procedures.\cite{Robin-1998}.  
The characteristic three-peak structure of the 
single-particle spectral features for ${\bf k} \geq {\bf k}_F$ in this part of the Brillouin 
zone around the hidden Fermi surface is a fingerprint of resonant pairing systems, 
highlighting their deviations from  BCS like superconductivity, for which the spectral features would 
invariably provide a two peak structure. The recent STM imaging studies are further proof of the non-BCS like features of the spectral properties attributed to in-gap single-particle contributions above $T_c$.

\section{Summary}
Exploiting the intrinsic metastability of hole doped cuprates, we investigated a scenario of resonant 
pairing, driven by dynamical local lattice instabilities. It invokes charge carriers, which 
simultaneously exist as itinerant quasi-particles and as localized self-trapped  bipolarons. 
The quantum fluctuations between the two configurations induce mutually (i)~a certain degree of 
itinerancy of the bare localized bipolarons and (ii)~localized features in the single-particle 
spectral function. This is evident in simultaneously featuring a low energy diffusive Bogoliubov 
branch below the chemical potential and single-particle excitations, accompanied by remnant bonding 
and anti-bonding states in the high energy response, above the chemical potential. These are 
finger prints of the non-BCS like physics of those cuprates, which should be interesting to test  experimentally and link up with studies of anomalous lattice properties which are at the origin of the dynamical lattice instability driven segregation of the homogeneous cuprate lattice structure into 
checker-board structures.

\section*{Acknowledgements}

I would like to thank my colleagues Tadek Domanski and Alfonso Romano  for having  participated in 
numerous theoretical studies on which this present attempt, to develop a global picture of the 
cuprates, is based. I thank Juergen Roehler, who kept me regularly informed about relevant 
experimental results as well and his critical and constructive comments, which helped me in my
understanding those materials.

\end{document}